\documentclass[aps,prd,twocolumn]{revtex4-1}

\usepackage{graphicx}  % needed for figures
\graphicspath{ {images/} } %the images are kept in a folder named images under the current directory.
\usepackage{latexsym}
\usepackage{slashed}
\usepackage{dcolumn}   % needed for some tables
\usepackage{bm}        % for math
\usepackage{amsmath}
\usepackage{amssymb}   % for math
\usepackage{longtable}
\usepackage{float}
\usepackage{array}
\newcommand{\be}{\begin{equation}}
\newcommand{\ee}{\end{equation}}
\newcommand{\bea}{\begin{eqnarray}}
\newcommand{\eea}{\end{eqnarray}}
\newcommand{\bma}{\begin{matrix}}
\newcommand{\ema}{\end{matrix}}

\newcommand{\bml}{\begin{mathletters}}
\newcommand{\eml}{\end{mathletters}}
\newcommand{\bes}{\begin{subequations}}
\newcommand{\ees}{\end{subequations}}

\newcommand{\bi}{\begin{itemize}}
\newcommand{\ei}{\end{itemize}}
\newcommand{\gev}{~{\rm GeV}}

\newcommand{\mev}{~{\rm MeV}}

\begin{document}
\title{Mirror fermions and the strong CP problem: A new axionless solution and experimental implications.}
\author{P. Q. Hung}
\email{pqh@virginia.edu}
\affiliation{Department of Physics, University of Virginia,
Charlottesville, VA 22904-4714, USA}

\date{\today}

\begin{abstract}
A new solution to the strong CP problem with distinct experimental signatures (long-lived particles) at the LHC is proposed. It is based on the Yukawa interactions between mirror quarks, Standard Model (SM) quarks and Higgs singlets. (Mirror quarks and leptons which include non-sterile right-handed neutrinos whose Majorana masses are proportional to the electroweak scale, form the basis of the EW-$\nu_R$ model.) The aforementioned Yukawa couplings can in general be complex and can contribute to $Arg\, Det M$ ($\bar{\theta} = \theta_{QCD} + Arg\, Det M$) at tree-level. The model contains a Peccei-Quinn-type global symmetry which allows it to rotate away $\theta_{QCD}$.The crux of matter in this manuscript is the fact that {\em no matter how large} the CP-violating phases in the Yukawa couplings might be, $Arg\, Det M$ can remain small i.e. $\bar{\theta} < 10^{-10}$ for reasonable values of the Yukawa couplings and, in fact, vanishes when the VEV of the Higgs singlet (responsible for the Dirac part of the neutrino mass in the seesaw mechanism) vanishes. The smallness of the contribution to $\bar{\theta}$ is {\em principally due} to the smallness of the ratio of the two mass scales in the seesaw mechanism: the Dirac and Majorana mass scales.
%Several prominent proposals have been put forward as solutions to the Strong CP problem, among which are the Axion model and models with new extra quarks.  

\end{abstract}

\maketitle

%\section{Introduction}
%This paper is an expanded version of a talk \cite{pqtalk} and a new take on the solution to the strong CP problem proposed in Ref.~(\cite{pqCP}).

It is a well-known fact that, although CPT appears to be respected as a symmetry of nature, CP and T are not as weak interaction experiments have shown us. Furthermore, studies of the QCD vacuum, the so-called $\theta$-vacuum, revealed that an additional CP-violating term is added to the Lagrangian in the form $\theta_{QCD} \, (g_{3}^2/32 \pi^2) G_{a}^{\mu \nu} \tilde{G_{\mu \nu}^{a}}$ \cite{thooft}. In addition, the electroweak sector contributes another similar term through quark mass matrices so that the total $\theta$ is now $\bar{\theta} = \theta_{QCD} + Arg Det M$.

Constraints coming from the absence of the neutron electric dipole moment give $\bar{\theta} < 10^{-10}$ \cite{theta}. This is the famous strong CP problem: why $\bar{\theta}$ should be so small. Several lines of approach toward a solution to the strong CP problem have been proposed. The most famous one is the Peccei-Quinn solution \cite{peccei} where a new chiral symmetry $U(1)_{PQ}$ was added. As has been noted by Ref.~\cite{jackiw}, a chiral rotation under $U(1)_{A}$ by an angle $\alpha$ changes the vacuum angle $\theta$ to $\theta + \alpha$. All vacuums are equivalent and one can {\em rotate} $\theta_{QCD}$ to zero as has been done in \cite{peccei}. Since one expects $U(1)_{PQ}$ to be spontaneously broken, this will induce an additional term $Arg Det M$ proportional to the vacuum expectation value (VEV) of a scalar field associated with $U(1)_{PQ}$. The axion, a pseudo Nambu-Goldstone boson introduced by Weinberg and Wilczek \cite{weinberg}, takes on an expectation value at the minimum of an effective scalar potential written under some approximation which participates in the cancellation of $\bar{\theta}$. This is, in a nutshell, the PQ solution to the strong CP problem. However, the axion is still elusive and its search is going on. 

A number of alternative, axionless models were constructed with varying degrees of assumptions such as soft CP breaking, P and T invariance or CP conservation of the Lagrangian \cite{mohapat}. In the latter class of models, CP is spontaneously broken giving rise to potential problems with issues such as domain walls.  
%A very early class of models solving the strong CP problem without the axion and using either soft CP breaking or simply P and T invariance can be found in \cite{mohapat}. Another line of approach \cite{SBCP}  is to assume CP conservation of the Lagrangian so that $\bar{\theta}=0$ at tree level and to postulate the existence of heavy fermions (within a Grand Unification context such as SU(5) or an extended gauge group $SM \times G$) to generate a non-vanishing $\bar{\theta}$ at loop levels. In this class of models, CP is spontaneously broken giving rise to potential problems with issues such as domain walls. 

In this paper, we propose a new solution to the strong CP problem which is based on the ingredients which are {\em already contained} in a model of non-sterile right-handed neutrinos with electroweak-scale masses: the EW-$\nu_R$ model \cite{nur}.
Our approach to the strong CP problem is similar in spirit to the PQ approach in that it also contains additional global symmetries, except for a few crucial differences. Our global symmetry which allows us to rotate $\theta_{QCD}$ away (as in the PQ approach) does not generate a dynamical axion-like field since we do not require the effective $\theta$ to be driven to zero. %s noticed by \cite{peccei}, there is no longer CP violation if the VEV of the aforementioned scalar goes to zero which is not expected. 
In our model, that VEV is proportional to the neutrino mass and the smallness of  $\bar{\theta}$ is linked to the smallness of the neutrino mass.

Three questions that need to be addressed are the following: 1) What chiral symmetry allows us to rotate $\theta_{QCD}$ to an equivalent vacuum where it is zero at tree level?; 2) Since CP can explicitly be violated by the complex Yukawa couplings, what prevents $Arg Det M$ from exceeding the upper bound of $10^{-10}$?; 3) Last but not least, can the solution be found {\em solely} within the gauge structure  of the SM, namely $SU(3) \times SU(2) \times U(1)$? 
%This paper will follow a two-step process. First, it will show that there is a chiral symmetry that allows $\theta_{QCD}$ to be rotated away. quivalent vacuum withSecond, it will show that $\bar{\theta}$ is small due to the fact that it is proportional to the neutrino mass.

To answer to question \#1, for completeness, a brief summary of a toy model presented in \cite{peccei} is in order and, in particular, the need to have a chiral symmetry to deal with $\theta_{QCD}$. First, the non-perturbative $\theta$-vacuum induces a term $\theta_{QCD} (g_{3}^2/32 \pi^2) \int d^4 x G_{a}^{\mu \nu} \tilde{G_{\mu \nu}^{a}}$ in the effective Lagrangian which violates CP \cite{thooft}. The toy model of \cite{peccei} simply consists of a single flavor of quark having a Yukawa coupling with a complex scalar of the form $\bar{\psi}_L G \phi \psi_R + \bar{\psi}_R G^{*} \phi^{*} \psi_L$. The Lagrangian of this toy model is invariant under a chiral rotation $\psi \rightarrow \exp (\imath \sigma \gamma_5) \psi$; $\phi \rightarrow \exp(-\imath   2 \sigma) \phi$. However, the associated chiral current is anomalous i.e. $\partial^{\mu} J_{5\mu} = (g_{3}^2/32 \pi^2) G_{a}^{\mu \nu} \tilde{G_{\mu \nu}^{a}}$. The change in the action is given by $\delta S = -\imath \int d^{4} x \partial^{\mu} J_{5\mu} \sigma= -\imath (2 \sigma) (g_{3}^2/32 \pi^2) \int d^4 x G_{a}^{\mu \nu} \tilde{G_{\mu \nu}^{a}}$. As stated by \cite{peccei}, this chiral rotation induces $\theta_{QCD} \rightarrow \theta_{QCD} - 2 \sigma$. In the parlance of Jackiw and Rebbi \cite{jackiw}, "all vacuua are degenerate in energy and define the same theory" and we can set $\theta_{QCD}$ to {\em zero}.

We will show that the EW-$\nu_R$ model \cite{nur} has the necessary ingredients such as the aforementioned chiral symmetry to solve the strong CP problem without the need of the axion.
%? The answer is yes. First, a quick tour of the model is in order here, to be followed by a description of the global symmetry relevant to the strong CP problem and the Yukawa interaction involved in the computation of $Arg Det M$.

What is the EW-$\nu_R$ model \cite{nur} and what has it accomplished? 1) It is based {\em solely} on the SM gauge group $SU(3)_C \times SU(2)_W \times U(1)_Y$. 2)  It contains mirror quarks and leptons out of those right-handed neutrinos emerge, are naturally non-sterile and have Majorana masses proportional to the electroweak scale $\Lambda_{EW} \sim 246 GeV$ and can be searched for at the LHC. This is the prime motivation of \cite{nur}: a direct test of the seesaw mechanism at colliders.
%The seesaw mechanism can be {\em tested} at colliders such as the LHC. 
3) It avoids the Nielsen-Ninomiya no-go theorem \cite{nielsen} (which states that one cannot put the SM on the lattice without having mirror fermions interacting with the same W and Z bosons) by postulating the very existence of these mirror fermions. Non-perturbative aspects of the SM such as the electroweak phase transition can now be studied on the lattice.4) It satisfies electroweak precision data represented by the parameters $S$, {$T$ and $U$ \cite{STU}. In fact, any BSM model is required to satisfy this first criterion \cite{tripletS}. 5) It accommodates the 125-GeV scalar \cite{125}. In particular, it provides two distinct scenarios for the 125-GeV scalar: {\em Dr Jekyll} (very SM-like) and {\em Mr Hyde} (very different from the SM Higgs), both of which give signal strengths consistent with experiment. 6) Analyses of productions and decays of mirror quarks and leptons have been performed with various constraints imposed on the model from existing experimental data. However, constraints coming from $\mu \rightarrow e \gamma$ and $\mu$ to $e$ conversion} generally requires $g_{Sl} < 10^{-4}$ ($g_{Sl}$ is the Yukawa coupling of the mirror lepton with the SM lepton and singlet Higgs) \cite{muegamma}. This gives rise to a characteristic decay signature: a displaced vertex. More details can be found in Ref.~\cite{muegamma,search}.
%and from $\bar{\theta}$ (strong CP)} requiring \textcolor{blue}{$g_{Sq}< g_{Sl}$}. This implies that characteristic signatures in the searches for mirror fermions will be \textcolor{blue}{{\em displaced vertices}}! 
 
%It predicts heavier scalars and pseudo-scalars.
%Last but not least, the part of the EW-$\nu_R$ model which is most relevant to the strong CP problem is the global symmetry imposed on the model in order to avoid terms which can spoil the seesaw mechanism. 
%Why is the aforementioned global symmetry required in our model?
Let us briefly recall that the Majorana mass for right-handed neutrinos comes from the term: $L_M =  g_M \, l^{M,T}_R \ \sigma_2  \tau_2  \tilde{\chi}  l^M_R$, where $l^M_R = (\nu_R, e^M_R )$ and $\tilde{\chi}$ is the Higgs triplet transforming as $(3,Y/2=1)$. This gives the Majorana mass term $M_R \nu^{T}_R \sigma_2 \nu_R$ where $M_R = g_M v_M \propto \Lambda_{EW} \sim 246 \gev$ with $\langle \tilde{\chi} \rangle = v_M$. 

The Dirac mass term comes from ${\mathcal L}_S = - g_{Sl} \,\bar{l}_L \, \phi_S \, l_{R}^M +  H.c.$ where $\phi_S$ is $SU(2)_W \times U(1)_Y$ singlet. This gives a Dirac mass $m_{\nu}^D = g_{Sl} \, v_S $ with $m_D \ll M_R$ \cite{nur}. The usual seesaw mechanism is then $m_\nu = m_D^2/M_R$ for the light neutrino and $M_R$ for the heavy right-handed neutrino. 

In order to prevent the appearance of terms such as $\bar{l}_L \tilde{\chi} l^{M}_R$ (A Dirac mass which is too big) and $l^{T}_L \ \sigma_2 \ \tau_2 \ \tilde{\chi} \ l_L$  which gives rise to an unwanted terms $M_L \nu_{L}^T \nu_L$ with $M_L \propto v_M$, a global symmetry was imposed: $U(1)_{MF}$ \cite{nur}. For symmetry reasons, this symmetry was generalized to $U(1)_{SM} \times U(1)_{MF}$ \cite{125}. A more complete realization of this global symmetry will be presented below. 

It is important to reemphasize that the aforementioned global symmetry is needed in the EW-$\nu_R$ model \cite{nur} in order to have right-handed neutrino masses to be proportional to the electroweak scale $\Lambda_{EW} \sim 246 GeV$ and to have the correct seesaw mechanism as described in \cite{nur}.

As emphasized in \cite{nur}, anomaly cancellation requires the existence of mirror quarks in addition to mirror leptons. This global symmetry applies to the quark sector as well. This is where the proposed solution to the strong CP problem comes in.

In order to illustrate the functionality of the  global symmetry $U(1)_{SM} \times U(1)_{MF}$ , we start out with the one-generation ( two flavors) case in the EW-$\nu_R$ model. 
%where there is no CKM but where CP violation is explicit in the Yukawa couplings between SM and mirror quarks. 
This helps to separate the two issues, that of the strong CP violation and that of the weak CP violation present in the CKM matrix for three (or more) generations of quarks.

 The gauge group is $SU(3)_C \times SU(2)_W \times U(1)_Y$. 
 We have one generation of SM quarks: $q_L= (u_L,d_L), u_R, d_R$, and one generation of mirror quarks: $q_R^M= (u_R^M, d_R^M), u_L^M, d_L^M$. (The full model can be found in the EW-$\nu_R$ model \cite{nur}. (For completeness, we also list the transformations of the leptons.) 
 
 The relevant scalar fields for the purpose of this manuscript are the doublets $\Phi^{SM}_{1} (Y/2=-1/2)$, $\Phi^{SM}_{2} (Y/2=+1/2)$, $\Phi^{M}_{1} (Y/2=-1/2)$, $\Phi^{M}_{1} (Y/2=+1/2)$ , and the complex singlet $\phi_S$. Here $Y/2=\pm1/2$ refers to the $U(1)_Y$ quantum numbers. We will focus on the quark sector from hereon. The rationale for the Higgs doublet content will be given below. Notice that the EW-$\nu_R$ model \cite{nur} also requires the existence of the triplets $\tilde{\chi}$ and $\xi$ but we will not discuss them here since they do not couple to the quarks.
 
 The Lagrangian of interest is given by
 \be
 \label{lagrangian}
 {\cal L}=  {\cal L}_{Kin} +  {\cal L}_{mass} +  {\cal L}_{mixing}  \, ,
 \ee
 where 
 \bea
 \label{mass}
 {\cal L}_{mass}& = & g_u \bar{q}_L  \Phi^{SM}_{1} u_R + g_d \bar{q}_L  \Phi^{SM}_{2} d_R  \nonumber \\
 &&+ g^{M}_{u} \bar{q}^M_R  \Phi^{M}_{1}  u^M_L+ g^{M}_{d} \bar{q}^M_R  \Phi^{M}_{2}  d^M_L  \nonumber \\
 &&+H.c. \,,
 \eea
 and
 \bea
 \label{mixing}
 {\cal L}_{mixing} &=& g_{Sq}\bar{q}_L \phi_S q^M_R + g_{Su}\bar{u}^M_L \phi_S  u_R  \nonumber \\
 &&+ g_{Sd}\bar{d}^M_L \phi_S  d_R + H.c. \,.
 \eea
% In Eq.~(\ref{mass}), $\tilde{\Phi}_{2}  \equiv \imath \tau_2 \Phi_{2}^*$ and $i=1,2$. A special case of Eq.~(\ref{mass}) was discussed in \cite{125}. In general, $a_{i}$, $b_i$, $c_i$ and $d_i$ are determined phenomenologically but it is outside the scope of this paper to discuss it here. In Eq.~(\ref{mixing}), $\phi_S$ is a complex Higgs singlet. 

${\cal L}$ is invariant under the following transformations of $U(1)_{SM} \times U(1)_{MF}$ 
\begin{eqnarray*}
\label{SMtransformation}
			U(1)_{SM}
			&:& q_L \;\rightarrow e^{-\imath \alpha_{SM}} \; q_L\,,\\
			&& (u_R, d_R) \;\rightarrow e^{\imath \alpha_{SM}} \; (u_R, d_R) \,, \\
			&& \Phi^{SM}_{1,2} \rightarrow  e^{-2\imath \alpha_{SM}} \Phi^{SM}_{1,2} \,.
			%&& e_R^{SM} \;\rightarrow e^{\imath \alpha_{SM}} \; e_R^{SM}
\end{eqnarray*} 
\begin{eqnarray*}
%\label{SMtransformation}
\label{MFtransformation}
			U(1)_{MF}
			&:& q_R^M \;\rightarrow e^{\imath \alpha_{MF}} \; q_R^M\,,\\
			&& (u_L^M, d_L^M) \;\rightarrow e^{-\imath \alpha_{MF}} \; (u_L^M, d_L^M) \,, \\
			&& \Phi^{MF}_{1,2} \rightarrow  e^{2\imath \alpha_{MF}} \Phi^{MF}_{1,2} \,,
			%&& e_R^{SM} \;\rightarrow e^{\imath \alpha_{SM}} \; e_R^{SM}
\end{eqnarray*} 
and
\be
\label{singlet}
\phi_S \rightarrow e^{-\imath(\alpha_{SM} + \alpha_{MF}}) \phi_S \,.
\ee

At this point, it is important to stress again the rationale for having $\Phi^{SM}_{1} (Y/2=-1/2)$, $\Phi^{SM}_{2} (Y/2=+1/2)$, $\Phi^{M}_{1} (Y/2=-1/2)$, $\Phi^{M}_{1} (Y/2=+1/2)$: They are needed in order for ${\cal L}$ to be invariant under $U(1)_{SM} \times U(1)_{MF}$ transformations. In essence, we are dealing here with a 2HDM (two Higgs doublet model) for the SM sector and another one for the mirror sector. This gives rise to a rich Higgs sector with interesting phenomenological implications \cite{higgs}.

%An important point is in order here. Notice that $U(1)_{SM,L} \times U(1)_{SM,R} \equiv U(1)_{SM,V} \times U(1)_{SM,A}$ and similarly for the mirror sector. The global symmetry can now be written as 
%\be
%U(1)_{SM,V} \times U(1)_{SM,A} \times U(1)_{MF,V} \times U(1)_{MF,A} \,.
%\ee
We shall be interested, in this paper, on the chiral symmetry $U(1)_{SM,A} \times  U(1)_{MF,A}$ which are contained in $U(1)_{SM} \times U(1)_{MF}$.  
%because of the anomaly.

 Under $U(1)_{SM,A} \times  U(1)_{MF,A}$, the SM and mirror quarks, $q$ and $q^M$, transform as
 %ymmetry $U(1)_{SM} \times U(1)_{MF}$, the SM and mirror fermions transform as
 \begin{eqnarray*}
 \label{SMtrans}
 &&q \rightarrow \exp(\imath \gamma_{5} \alpha_{SM}) q \,, \\
 &&q^{M} \rightarrow \exp(\imath \gamma_{5} \alpha_{MF}) q^{M} \,.
\end{eqnarray*} 
 
% It is straightforward to see that ${\cal L}_{Kin}$ is invariant under the transformations (\ref{SMtrans}) and (\ref{MFtrans}). As for ${\cal L}_{mass}$, this is similar to the QCD Lagrangian where the mass terms break explicitly chiral symmetry. Here, because of the presence in (\ref{mass}) of both $\Phi_{2}$ and $\tilde{\Phi}_{2}  \equiv \imath \tau_2 \Phi_{2}^*$, one cannot find a transformation for $\Phi_{2}$ to make ${\cal L}_{mass}$ invariant under (\ref{SMtrans}) and (\ref{MFtrans}). In fact, as stated in \cite{nur}, $\Phi_{2}$ carries no quantum number of the aforementioned global symmetry. ${\cal L}_{mass}$ represents the explicit breaking of chiral symmetry similar to the usual QCD Lagrangian. According to \cite{nur}, $\Phi_{2}^{i} \rightarrow \Phi_{2}^{i}$ under the above global symmetry.
 
 We now turn to ${\cal L}_{mixing}$ and discuss the implication of $U(1)_{SM} \times U(1)_{MF}$ on $\theta_{QCD}$.

Notice that $g_u$, $g_d$, $g^{M}_{u}$, $g^{M}_{d}$, $g_{Sq}$, $g_{Su}$ and $g_{Sd}$ can, in general be complex. If we absorb the phases into $u_R$, $u^{M}_L$, $d_R$ and $d^{M}_L$ to make the {\em diagonal} elements of the ($2 \times 2$) up and down mass matrices {\em real} then the {\em off-diagonal} elements stay {\em complex}.
 Furthermore, the global symmetry $U(1)_{SM} \times U(1)_{MF}$  was invoked in \cite{nur,125} to ensure that the Yukawa couplings take the form as shown in Eq.~(\ref{mass}). 
 
 $\langle \Phi^{SM}_{1,2}  \rangle = v_{1,2}$ and $\langle \Phi^{M}_{1,2}  \rangle = v^{M}_{1,2}$ give non-vanishing masses to the SM and mirror quarks, namely $m_u$, $m_d$, $M_u$ and $M_d$ respectively. (From \cite{nur,125}, $(\sum_{i=1,2} v_i^2 + v^{M,2}_{i}) + 8 v_M^2 = (246 \gev)^2$ where $v_M$ is the VEV of the Higgs triplet which gives the Majorana mass to $\nu_R$.)  The mass mixing between SM and mirror quarks comes from Eq.~(\ref{mixing}). Writing $g_{Sq} = |g_{Sq} | \exp(\imath \theta_q)$, $g_{Su} = |g_{Su} | \exp(\imath \theta_u)$ and $g_{Sd} = |g_{Sd} | \exp(\imath \theta_d)$ and with $\langle \phi_S \rangle = v_S$, we obtain the following mass matrices ($m_u$, $M_u$, $m_d$ and $M_d$ are real)
 
 \begin{equation}
	\label{Mu}
	{\cal M}_u=\left( \begin{array}{cc}
	m_u & |g_{Sq} | v_S \exp(\imath \theta_q) \\
	 |g_{Su} | v_S  \exp(\imath \theta_u)& M_u
	\end{array} \right) \,,
\end{equation}
 \begin{equation}
	\label{Md}
	{\cal M}_d=\left( \begin{array}{cc}
	m_d& |g_{Sq} | v_S \exp(\imath \theta_q) \\
	 |g_{Sd} | v_S  \exp(\imath \theta_d)& M_d
	\end{array} \right) \,.
\end{equation}

It is important  to emphasize again a point that has been made above concerning the Dirac part of the neutrino mass, namely $m_{\nu}^D = |g_{Sl}| \, v_S \ll M_R \sim \Lambda_{EW}$. If we assume $|g_{Su},|g_{Sd} | \leq |g_{Sl} |$ (as shown below) which implies $ (|g_{Su},|g_{Sd} |)\, v_S \ll m_{u,d}, M_{u,d}$, it can then be seen that the complex off-diagonal elements of ${\cal M}_u$ and ${\cal M}_d$ are {\em small perturbations} to the diagonal (real) elements.

Two important observations are in order here.

\bi

\item The case where $\langle \phi_S \rangle = 0$.

From Eqs.~(\ref{Mu},\ref{Md}), one can see that ${\cal M}_u$ and ${\cal M}_d$ are {\em real} and {\em diagonal}. Under a chiral transformation
\bea
\label{chiral}
(u,d)^{SM} &\rightarrow& \exp(\imath \alpha_{SM} \gamma_5) (u,d)^{SM} ; \\   
(u,d)^M &\rightarrow& \exp(\imath \alpha_{MF} \gamma_5) (u,d)^M ; \\
\phi_S &\rightarrow& \exp(- \imath (\alpha_{SM}+ \alpha_{MF})) \phi_S \, ,
\eea
the vacuum angle $\theta_{QCD}$ is changed as follows
\be
\label{change}
\theta_{QCD} \rightarrow \theta_{QCD} - (\alpha_{SM} + \alpha_{MF}) \,.
\ee
Since ${\cal M}_u$ and ${\cal M}_d$ are {\em real} and {\em diagonal}, there is {\em no additional phase} from the mass matrices i.e. $Arg Det ({\cal M}_u {\cal M}_d) =0$. All vacuua are equivalent and we can choose the CP-conserving one
\be
\label{condition}
 \theta_{QCD} - (\alpha_{SM} + \alpha_{MF}) =0 \,,
\ee
for arbitrary $\theta_{QCD}$, $\alpha_{SM} $ and $\alpha_{MF}$. In the absence of mixing in (\ref{Mu}) and (\ref{Md}), there is {\em no strong CP violation}. The above results come from the following considerations of chiral currents.

The chiral currents corresponding to SM and mirror quarks are anomalous
\be
\label{SMchiral}
\partial^{\mu} J^{SM}_{5\mu} = (g_{3}^2/16 \pi^2) G_{a}^{\mu \nu} \tilde{G_{\mu \nu}^{a}} \,,
\ee
\be
\label{MFchiral}
\partial^{\mu} J^{MF}_{5\mu} = (g_{3}^2/16 \pi^2) G_{a}^{\mu \nu} \tilde{G_{\mu \nu}^{a}} \,,
\ee
where $SM$ and $MF$ stand for Standard Model and Mirror Fermion respectively. (Also, let us remind ourselves that one has the factor $1/16$ instead of $1/32$ in (\ref{SMchiral}) and (\ref{MFchiral}) because we have two flavors in this example.) From Eq.~(\ref{SMchiral},\ref{MFchiral}), one obtains the following anomalous and non-anomalous combinations: $J^{SM}_{5\mu} + J^{MF}_{5\mu}$ and $J^{SM}_{5\mu} - J^{MF}_{5\mu}$ respectively
\be
\label{anomalous}
\partial^{\mu} (J^{SM}_{5\mu} + J^{MF}_{5\mu})= (g_{3}^2/16 \pi^2) G_{a}^{\mu \nu} \tilde{G_{\mu \nu}^{a}} \,,
\ee
\be 
\label{nonanomalous}
\partial^{\mu} (J^{SM}_{5\mu} - J^{MF}_{5\mu})= 0 \,.
\ee
Notice that Eqs.~(\ref{anomalous},\ref{nonanomalous}) are reminiscent of similar ones concerning the SM model where, at the quantum level, the $B+L$ current is anomalous while the $B-L$ one is conserved. They are interesting in their own right but it is beyond the scope of this paper to go beyond the strong CP problem. The change in the action takes the form
\bea
\label{action}
\delta S &=& -\imath \int d^{4} x \partial^{\mu} (J^{SM}_{5\mu} \, \alpha_{SM} + J^{MF}_{5\mu} \, \alpha_{MF})/2 \\
              &=& -\imath (\alpha_{SM}+ \alpha_{MF}) (g_{3}^2/32 \pi^2) \int d^4 x G_{a}^{\mu \nu} \tilde{G_{\mu \nu}^{a}} \,.
\eea
This implies (\ref{change}).

Having fixed the CP-conserving vacuum (\ref{condition}), let us move on to the case where $\langle \phi_S \rangle = v_S \neq 0$ which is the desired feature of the EW-$\nu_R$ model \cite{nur}. 
%One can treat this as a {\em small} perturbation to the strong CP-conservation situation which comes about because of the presence of the chiral symmetry described above.

\item The case where $\langle \phi_S \rangle = v_S \neq 0$.

As we have seen above, this is the situation in which ${\cal M}_u$ and ${\cal M}_d$ acquire complex off-diagonal elements. The diagonalization of these two matrices introduce the well-known additional phase to the $\theta$ vacuum: $Arg Det ({\cal M}_u {\cal M}_d) \neq 0$. 
%As we have discussed above, the off-diagonal elements can be considered to be small "perturbations" of matrices with real diagonal elements. 
The question now is how much deviation one obtains in moving from a CP-conserving vacuum $\bar{\theta} = 0$ to
a CP-violating vacuum
\be
\label{thetabar}
\bar{\theta} \equiv \theta_{Weak} = Arg Det ({\cal M}_u {\cal M}_d) \,.
\ee
where $\theta_{Weak}$ stands for the contribution to the $\theta$ vacuum coming from the quark mass matrices. 

%From hereon, we will call $\bar{\theta}$ by $\theta_{Weak}$.

Finally, an important point is in order here. When $\phi_S$ develops a VEV $\langle \phi_S \rangle = v_S$, not only does it give the Dirac mass term in the seesaw mechanism of \cite{nur}, it also spontaneously breaks $U(1)_{SM} \times U(1)_{MF}$ down to $U(1)_{(SM)-(MF)}$. With $\phi_S= Re\phi_S + \imath Im \,\phi_S$, one would, in principle, expect $Im\,\phi_S$ to become a Nambu-Goldstone (NG) boson. However, as emphasized in \cite{nur}, the important point found in \cite{chanowitz} is that, for a proper vacuum alignment, the Higgs potential contains a term that includes a mixing between the two Higgs triplets $\tilde{\chi}$ and $\xi$ of the form $\lambda_4 (\xi^0 \, \tilde{\chi}^0+...)$. (In \cite{125}), $\lambda_4 \rightarrow \lambda_5$.) A term like $\xi^0 \, \tilde{\chi}^0$ explicitly breaks $U(1)_{SM} \times U(1)_{MF}$ since under which $\xi^0$ is a singlet while $\tilde{\chi}^0$ is not. One would then expect $Im\,\phi_S$ to become a pseudo-Nambu-Goldstone (PNG) boson (denoted by $A_S$ from here on) whose mass would be proportional to $\lambda_4$. If $\phi_S$ were NOT present as a fundamental scalar like in the model of \cite{nur} the SSB of $U(1)_{SM} \times U(1)_{MF}$ would be accomplished by  a QCD condensate of the form $\langle \bar{q}_L q^{M}_R \rangle$ with a scale around a $\gev$. Notice the PNG in this case would be a $\bar{q}_L q^{M}_R$ bound state and NOT $\bar{q}_L u_R$ for example.  The presence of $\phi_S$ in the model of \cite{nur} removes this possibility. Although it is an interesting topic in its own right, it is beyond the scope of this paper to discuss it further. The possibly interesting phenomenology of the PNG $A_S$ will treated elsewhere.

\ei

We now move on to the discussion of the size of $Arg Det ({\cal M}_u {\cal M}_d)$.

From Eq.~(\ref{Mu},\ref{Md}), we obtain ($C_u \equiv m_u M_u$, $C_d \equiv m_d M_d$, $C_{Su} \equiv  |g_{Sq}||g_{Su}| v_S^2$ and $C_{Sd} \equiv  |g_{Sq}||g_{Sd}| v_S^2$)
\bea
Arg Det ({\cal M}_u {\cal M}_d)&=&Arg \{ (C_u - C_{Su} \exp[\imath(\theta_q + \theta_u)]) \nonumber \\
&&(C_d - C_{Sd} \exp[\imath(\theta_q + \theta_d)])\} \,.
\eea
%Neglecting the term proportional to $C_{Su} C_{Sd}$ since (as we shall explain below) $C_{Su} C_{Sd} \ll C_u C_d, C_{Su}C_d, C_{Sd} C_u$ 
With $r_u = C_{Su}/C_u=(|g_{Sq}||g_{Su}|/g_{Sl}^2)/(m_u M_u)$, $r_d= C_{Sd}/C_d=(|g_{Sq}||g_{Sd}|/g_{Sl}^2)/(m_d M_d)$,and taking into account the fact that $m_D^2 \ll (m_{u,d} M_{u,d}$ and consequently $r_{u,d} \ll 1$, $\theta_{Weak} \equiv Arg Det ({\cal M}_u {\cal M}_d)$ takes the form
%\be
%\label{thetaw1}
%\theta_{Weak} \approx \tan^{-1} \frac{-(C_{Su} C_{d}\sin(\theta_q + \theta_u) + C_{Sd} C_{u}\sin(\theta_q + \theta_u))}{C_d C_u -C_{Su} C_{d}\cos(\theta_q + \theta_u)- C_{Sd} C_{u}\cos(\theta_q + \theta_u)}
%\ee
%Defining
%\be
%%\label{ru}
%r_u = \frac{C_{Su}}{C_u}=\frac{|g_{Sq}||g_{Su}| v_S^2}{m_u M_u} \,,
%\ee
%\be
%\label{rd}
%r_d= \frac{C_{Sd}}{C_d}=\frac{|g_{Sq}||g_{Sd}| v_S^2}{m_d M_d} \,,
%\ee
%Eq.~(\ref{thetaw1}) can now be put in a neater form
\be
\label{thetaw2}
\theta_{Weak} \approx -(r_u \sin(\theta_q + \theta_u) + r_d \sin(\theta_q + \theta_d))
\ee 

%We are now ready to discuss the implications of Eq.~(\ref{thetaw2}).
The above results can be generalized to the three-family case with little changes in the conclusion. It is beyond the scope of this short letter to present it here.
\bi
\item First, we notice from Eq.~(\ref{thetaw2}) that $\theta_{Weak} =0$ when the VEV of the singlet Higgs vanishes i.e. when $v_S=0$. This is valid for {\em any} value of the phases $\theta_{q,u,d}$.

\item $\theta_{Weak}$ can also vanish if all the phase angles vanish or if $\theta_q =-\theta_u=-\theta_d$. Since theses are special cases, we will not consider them here but will instead keep them arbitrary.

\item As shown in \cite{nur}, a non-vanishing value for $v_S$ implies a non-vanishing Dirac mass of the neutrino participating in the seesaw mechanism i.e. $m_{\nu} = m_D^2/M_R$. From \cite{nur}, $m_D = g_{Sl} v_S$ coming from the interaction $g_{Sl} \bar{l}_L \phi_S \l^M_R + H.c.$ where $l_L=(\nu_L, e_L)$ and $l^M_R =(\nu_R, e^M_R)$. Here $M_R$ is the Majorana mass of the right-handed neutrino coming from $g_M \, (l^{M,T}_R \ \sigma_2)(\imath  \tau_2) \ \tilde{\chi} \ l^M_R$ where $\chi$ is a triplet Higgs with $Y/2=1$ and whose VEV is $v_M$.

\item Since $M_R > M_Z/2 \sim 45 \gev$ (from the Z-width constraint), one gets $m_D < 100\, keV$ \cite{nur}. 

%\item One can rewrite $r_u$ and $r_d$ as
%\be
%\label{newru}
%r_u=(\frac{|g_{Sq}||g_{Su}|}{g_{Sl}^2})(\frac{m_D^2}{m_u M_u}) \,,
%\ee
%\be
%\label{newrd}
%r_d=(\frac{|g_{Sq}||g_{Sd}|}{g_{Sl}^2})(\frac{m_D^2}{m_d M_d}) \,.
%\ee
 %$r_{u,d} \ll 1$ if one assumes that $|g_{Sq}||g_{Su,Sd}| \leq g_{Sl}^2$.

%\item It is interesting to notice that, since the Majorana mass of the right-handed neutrinos is also proportional to the electroweak scale $246 \gev$, $r_u$ and $r_d$ which will determine the size of $\theta_{Weak}$ have the following proportionality
%$r_u \propto m_\nu / m_u; r_d \propto m_\nu / m_d$
%and vanish as $m_\nu \rightarrow 0$.
%\item One can now rewrite $\theta_{Weak}$ as
%\be
%\label{thetaw3} 
%\theta_{Weak} \approx -(r_u \sin(\theta_q + \theta_u) + r_d \sin(\theta_q + \theta_d)) \,.
%\ee

\item As discussed in \cite{125}, one expects the mirror quarks to be heavy. For the sake of estimation, we shall take $M_u \sim M_d \sim 400 \gev$. Furthermore, since we are dealing with the one-generation case, let us take the most extreme case, namely $m_u \sim 2.3 \mev$ and $m_d \sim 4 \mev$. With the constraint $m_D < 100\, keV$, one obtains the following bound
\bea
\label{bound}
\theta_{Weak} &<& -10^{-8}\{(\frac{|g_{Sq}||g_{Su}|}{g_{Sl}^2})\sin(\theta_q + \theta_u)  \nonumber \\
&&+ (\frac{|g_{Sq}||g_{Sd}|}{g_{Sl}^2})\sin(\theta_q + \theta_d)\}
\eea
\item What does the inequality (\ref{bound}) imply? $|\theta_{Weak}| < 10^{-10}$ regardless of the values of the CP phases.
Even if one had {\em maximal} CP violation in the sense that $\theta_q + \theta_u \sim \theta_q + \theta_d \sim \pi/2$, $|\theta_{Weak}| < 10^{-10}$ provided $|g_{Sq}| \sim |g_{Su}| \sim |g_{Sd}| \sim 0.1 g_{Sl}$. 

\item This has interesting phenomenological implications concerning the searches for mirror quarks and leptons at the LHC \cite{search}. In fact, constraints coming from $\mu \rightarrow e \gamma$, $\mu$-$e$ conversion and the electron electric dipole moment \cite{muegamma} indicate that $g_{Sl} < 10^{-4}$ which would imply in the present context that $|g_{Sq}| \sim |g_{Su}| \sim |g_{Sd}| < 10^{-5}$. This implies the possibility of observing the decays of mirror quarks and leptons from the process $f^M \rightarrow f + \phi_S$ (where $f^M$ and $f$ stand for mirror and SM fermions respectively) at {\em displaced vertices} (large decay lengths) because of the small Yukawa couplings.
 
\item As already pointed out in \cite{nur}, the mass mixing between SM and mirror quarks is tiny, being proportional to the ratio of neutrino to quark mass. For most practical purpose, the mass eigenstates are approximately pure SM and mirror states.
\ei

A full analysis will involve three generations and will be more complicated. As opposed to the one-generation case where we have a $2 \times 2$ matrix, we will now have a $6 \times 6$ matrix of the form
\begin{equation}
	\label{6x6u}
	{\cal M}_u=\left( \begin{array}{cc}
	M_u & M_{q_L q^M_R} \\
	 M_{u_R u^{M}_L}& M_{u^M}
	\end{array} \right) \,,
\end{equation}
\begin{equation}
	\label{6x6d}
	{\cal M}_d=\left( \begin{array}{cc}
	M_d & M_{q_L q^M_R} \\
	 M_{d_R d^{M}_L}& M_{d^M}
	\end{array} \right) \,,
\end{equation}
where each element of the above matrices are $3 \times 3$ matrices. The matrices $M_{q_L q^M_R}$, $M_{u_R u^{M}_L}$ and $M_{d_R d^{M}_L}$ contain matrix elements which are proportional to the VEV of the singlet Higgs field, namely $v_S$. As we have shown above for the one-generation case, these are much smaller than matrix elements of $M_u$, $M_d$, $M_{u^M}$ and $M_{d^M}$. For this reason, those mass matrices can be diagonalized separately, neglecting mixing. Furthermore, we believe that the result for $\theta_{Weak}$ will not be too different for that given in Eq.~(\ref{bound}).

We carried out an analysis based on a simplified version of the full model. (The full analysis is beyond the scope of the paper and will be presented elsewhere.) We assume that $M_{u^M}$ and $M_{d^M}$ are diagonal. The problem is now reduced to a diagonalization of a $4 \times 4$ matrix of the form
\begin{equation}
	\label{4x4u}
	\tilde{{\cal M}}_{u,k}=\left( \begin{array}{cc}
	M_u & M^{i4}_{q_L q^M_R} \\
	 M^{4j}_{u_R u^{M}_L}& m_{u^M,k}
	\end{array} \right) \,,
\end{equation}
where $i,j,k=1,2,3$ and where $m_{u^M,k}$ denotes the mass of th $k$th up mirror quark. Similarly, one has
\begin{equation}
	\label{4x4d}
	\tilde{{\cal M}}_{d,k}=\left( \begin{array}{cc}
	M_d & M^{i4}_{q_L q^M_R} \\
	 M^{4j}_{d_R d^{M}_L}& m_{d^M,k}
	\end{array} \right) \,.
\end{equation}

For simplicity, let us assume $m_{u^M,k}= m_{u^M}$ and $m_{d^M,k}=m_{d^M}$. In a recent work, Ref.~\cite{quarkmass} constructs phenomenogically the up and down-quark mass matrices $M_u$ and $M_d$. 
%which can reproduce the known phenomenology of the CKM matrix and quark masses. 
These matrices turn out to be Hermitian and have real determinants. A simple calculation shows that the results are very similar to the one-generation case with similar quantities such as $r_u$ and $r_d$. Once again, we find $\theta_{Weak} \propto m_{\nu}/m_{u},m_{\nu}/m_{d}$.

%with the same conclusion as the one obtained for the one-generation case.
We conclude by summarizing the salient points of this paper.

1) CP violation in the strong sector {\em does not need to vanish}! This violation is {\em tiny} because it is linked to the neutrino mass. As such, there is no need for the presence of the axion. This axionless solution is based on the EW $\nu_R$ model \cite{nur}.

2) The EW $\nu_R$ model \cite{nur} was first conceived to provide a testable model of the seesaw mechanism by making right-handed neutrinos non-sterile and sufficiently "light" (i.e. with a mass $M_R$ proportional to the electroweak scale). These right-handed neutrinos do not come by themselves but are members of right-handed $SU(2)_W$ doublets which include right-handed mirror leptons. $SU(2)_W$ anomaly freedom dictates that one should also have doublets of right-handed mirror quarks. (The model includes per family $SU(2)_W$- singlets: $e_R$, $u_R$, $d_R$ for the SM fermions and $e^M_L$, $u^M_L$ and $d^M_L$ for the mirror fermions.) In fact, the SM and mirror sectors allow us to evade the Nielsen-Ninomiya no-go theorem \cite{nielsen} which forbids the chiral SM model to be put on the lattice, a special feature with possible interesting consequences.

3) Ingredients contained in the EW $\nu_R$ model are precisely those that allow us to solve the strong CP problem. First, the chiral symmetries that allow us to rotate the $\theta$ angle to zero have been presented above. Second, by mixing the left-handed SM lepton doublets with the right-handed mirror lepton doublets through the Higgs singlet fields, one obtains the neutrino Dirac mass $m_D$ which participates in the seesaw mechanism ($m_D^2/M_R$). This same mixing also operates in the quark sector giving rise to mixing between SM and mirror quarks in the mases matrices, which, in turn, contributes to the CP-violating parameter $Arg Det (M_u M_d)$ in an interesting way. It vanishes if $m_D$ goes to zero and is small ($< 10^{-10}$) because $m_D \ll M_R$ as in the seesaw mechanism. It is surprising that two seemingly unrelated phenomena find a common niche in the EW $\nu_ toR$ model. 

4) The constraint coming from $\bar{\theta}$ turns into a constraint on the aforementioned Yukawa couplings $g_{Sq}$ (in conjunction with constraints on $g_{Sl} < 10^{-4}$) and leads to distinct experimental signatures of the model: the decays of mirror quarks and leptons occur at {\em displaced vertices}. In fact, there is presently an initiative under the name "The LHC LLP Community" (with LLP standing for long-lived particles) with the aim of providing a space for experimentalists and theorists to discuss search strategies for long-lived particles whose possible existence might have been so far overlooked \cite{beacham}.

As mentioned above, there is a rich Higgs sector to be explored phenomenologically and experimentally in order to probe the nature of the electroweak symmetry breaking \cite{higgs}.

5) It would be interesting to coordinate the search for the neutron electric dipole moment with that for the absolute mass of the neutrino.

I would like to thank Alfredo Aranda and T.c.Yuan for useful remarks.
%
%
%%%%%%%%%%%%%%%%%%%%
\appendix
%%%%%%%%%%%%%%%%%%%%
%
%
%%%%%%%%%%%%%%%%%%%%	
%%%%%%%%%%%%%%%%%%%%
%\FloatBarrier
%%%%%%%%%%%%%%%%%%%%
%
%
%
%%%%%%%%%%%%%%%%%%%%
%

\end{document}